\documentstyle[aps,amsmath,amssymb]{revtex}
\tightenlines
\begin{document}

\title{Sub and Super-Luminal Propagation of Intense Pulses in
Media with Saturated and Reverse Absorption}
\author{G.S.Agarwal, and Tarak Nath Dey}
\address{Physical Research Laboratory, Navrangpura, Ahmedabad-380 009, India}
\date{\today}
\maketitle
\begin{abstract}
We develop models for the propagation of intense pulses in solid
state media which can have either saturated absorption or exhibit
reverse absorption . We show that the experiments of Bigelow {\it
et al.}[Phys. Rev. Lett. {\bf 90}, 113903 (2003); Science {\bf
301}, 200 (2003).] on subluminal propagation in Ruby, and
superluminal propagation in Alexandrite are well explained by
modelling them as three level and four level systems respectively,
coupled to Maxwell equations. We present results well beyond the
traditional pump-probe approach.

\end{abstract}

Since the discovery of ultraslow light with a group velocity 17
m/sec in a Bose condensate by Hau {\it et
al.}\cite{Hau,Kasapi,Mastko} many experiments have reported slow
light in a variety of media \cite{Kash,Bud,Turukhin}. Kash {\it et
al.} \cite{Kash} demonstrated light propagation with a group
velocity of 90 m/sec at room temperature in Rb vapor. Using Zeeman
coherences, Budker {\it et al.} \cite{Bud} reported slow light
with group velocity 8 m/sec in Rb vapor. Hemmer {\it et al.}
\cite{Turukhin} first reported slow light in solid state material
namely Pr doped Y$_2$SiO$_5$, maintained at a cryogenic
temperature of 5K. The slow light ideas have been successfully
used in storage and retrieval of light pulses \cite{Liu,Phillips}.
The understanding of storage and retrieval of light pulses has
been provided by Dey and Agarwal \cite{Tarak}, using the adiabaton
theory of Grobe, Hioe and Eberly \cite{Grobe}.

Work on pulse propagation continues to produce interesting
results\cite{Cerboneschi,Eberly,Juzeliunas,Mewes}. Recently,
Bigelow {\it et al.} \cite{Bigelow} showed the propagation of
light pulses in Ruby at a group velocity of 57.5 m/sec. This
experiment differs considerably from all earlier experiments which
were based on electromagnetically induced transparency
\cite{Harris}. Bigelow {\it et al.} recognize that a two level
system driven by a strong field and a probe gives rise to a hole
in the probe response function with a width of the order of
$1/T_1$, where $T_1$ is the  longitudinal relaxation time
\cite{Hillman}. Note that in a material like Ruby the transverse
relaxation time $T_2 \ll T_1$. These authors also discovered that
they need not use separate pump and probe fields. A field with
peak power of the order of saturation intensity could be slowed
down considerably to about 57.5 m/sec. Further Bigelow {\it et
al.} \cite{Boyd} extended their work to a material like
Alexandrite, where they reported superluminal propagation. There
have been several earlier reports of superluminal propagation in
solid state systems \cite{Chu}; and in vapors \cite{Wang,Chiao}.

The purpose of this letter is to study the propagation of intense
pulses in a homogeneously broadened medium, like Ruby, which can
exhibit saturated absorption or a medium like Alexandrite, which
can exhibit reverse absorption. Note that the pulse propagation in
a nonlinear transparent medium has been extensively
studied\cite{Allen}. The systems studied here differ from the
previous studies as our systems posses very strong transverse and
longitudinal relaxation effects. In order to model the
experiments, we model Ruby as a three level system and Alexandrite
as a four level system. We solve the system of coupled equations
numerically to delineate various aspects of pulse propagation. We
do not make any approximation on the strength of the pulses so
that we can model experimental observations on strong pulses. We
calculate group velocity from the relative delay or advancement
between the reference pulse and the output pulse. We present
numerical results on the propagation of Gaussian and modulated
pulses and show good agreement with the experimental data of
Refs.\cite{Bigelow,Boyd}. The experiments of Bigelow {\it et al.}
fall in two categories.  One consists of a weak probe and a strong
coherent cw field. These were explained in terms of the response
to a probe field of a two level medium pumped by a coherent
field\cite{Mollow}. The other category consists of the self delay
of a strong pulse. The latter requires solutions of the coupled
Maxwell-Bloch equations and this is the case we concentrate on.

{\it Media with saturated absorption}----For pulse propagation in
the three level model in Ruby, we represent the ground state as
$|g\rangle$, the $^4F_2$ absorption band as $|e_1\rangle$ and the
levels $2\bar{A}$ and $\bar{E}$ as $|e_2\rangle$. In Ruby one has
very rapid decay of the level $|e_1\rangle$ to $|e_2\rangle$, and
as a result some of the coherences become irrelevant on
experimental time scale. The density matrix equations for the
model of the Fig. 1 are
\begin{eqnarray}
 \dot{\rho}_{_{gg}} &=& 2\Gamma_2\rho_{_{_{22}}} + i
 \Omega(\rho_{_{_{1g}}}-\rho_{_{_{g1}}})\nonumber\\
 \dot{\rho}_{_{_{22}}} &=&
 2\Gamma_1\rho_{_{_{11}}}
 -2\Gamma_2\rho_{_{_{22}}}\nonumber\\
 \dot{\rho}_{_{_{1g}}} &=& -\Gamma_1\rho_{_{_{1g}}} + i
 \Omega(\rho_{_{gg}}-\rho_{_{_{11}}})\nonumber\\
 \rho_{_{gg}}&+&\rho_{_{_{11}}}+\rho_{_{_{22}}}=1,
 \end{eqnarray}
  where
  $\rho_{_{_{ij}}}=\langle e_{_{i}}|\rho|e_{_{j}}\rangle$,
 $i,j=1,2$.
 The Rabi frequency $2\Omega$ is defined by
 $2\Omega(z,t)=2\vec{d}_{_{1g}}~\cdot~{\cal\vec{E}}(z,t)/\hbar$, where
 $\vec{d}_{_{1g}}$ is the dipole matrix element and
 ${\cal\vec{E}}(z,t)$ is the envelop of the pulse.
 We assume that the carrier frequency, $\omega$, is on resonance with the frequency of
 the $|e_1\rangle\longleftrightarrow|g\rangle$ transition. Under the
 approximations, $\Gamma_1\gg\Gamma_2, \Omega$;~
 $\dot{\rho}_{_{1g}}\sim 0$, we derive the approximate equation for
 the evolution of the ground state population as
 \begin{equation}
 \dot{\rho}_{gg}~=~2\Gamma_2(1-\rho_{gg})
 ~-~2\frac{\Omega^2}{\Gamma_1}\rho_{gg}
 \end{equation}
 Note that we can prove that $\dot{\rho}_{_{11}}\approx 0$, if
 $\Gamma_1\gg\Gamma_2, \Omega$. Under the same conditions and the
 slowly varying envelop approximation, the evolution
 equation for the Rabi frequency of the field is governed by
 \begin{equation}
 \frac{\partial \tilde{\Omega}}{\partial z}= -
 \frac{\alpha_0}{2}\tilde{\Omega}\rho_{gg},~~~\tilde{\Omega}=\Omega/\Omega_{sat},
 \end{equation}
 where $\alpha_0 =4\pi\omega|d_{_{1g}}|^2/c\hslash\Gamma_1$ and  $\Omega_{sat}=2\sqrt{\Gamma_1
 \Gamma_2}$. In Eqs. (2) and (3) we have used the pulse coordinates i.e,
 $t-z/c, z$. The time derivative in Eq. (2) is with respect to
 $(t-z/c)$. The time $t$ can be
 expressed in units of $1/2\Gamma_2$. For numerical computation, we
 consider two types of input pulses, viz, a Gaussian pulse with a
 temporal width $\gtrsim 1/\Gamma_2$
 \begin{equation}
 \tilde{\Omega}_{in}=\tilde{\Omega}^0~e^{[-t^2/2\sigma^2]}
 \end{equation}
 and amplitude modulated pulse
 \begin{equation}
 \tilde{\Omega}^2_{in}(t)=I=I_0(1+m\cos[\Delta t]).
 \end{equation}
 The Equations (2) - (5) are our working equations. We use these
 for numerical computations. We calculate the evolution of the
 pulse for arbitrary values of $\tilde{\Omega}^0$ or $I_0$. Some
 typical results for the Gaussian pulses are shown in the
 Fig. 2.
 We get group velocities in the range 50 m/sec for
 $\Omega/\Omega_{sat}\backsim1$ and the transmission is rather
 small. In Fig. 3, we exhibit the behavior of $v_g$ and
 transmission as a function of the input intensity. These results
 are in agreement with the experimental findings of transmission
 in the range $0.1\%$.
 In Fig. 3, we also show for comparison the
 results of the group velocity and the transmission for the
 propagation of an intense pulse through a two level
 system described by the traditional Bloch equations.
 The coupled Maxwell-Bloch equations under the approximation
 $ T_1 \gg T_2 $ are given by
 \begin{eqnarray}
 \dot{\rho}_{_{11}}T_{_{1}}&=&-\rho_{_{11}}~+~2|\tilde {\Omega}|^2~(1 -
 2\rho_{_{11}})\\
 \rho_{_{gg}}&+&\rho_{_{11}}=1\\
 \frac{\partial \tilde{\Omega}}{\partial z}&=& -
 \frac{\alpha_0}{2}\tilde{\Omega}(1-2\rho_{_{11}}),
 \end{eqnarray}
 where $\tilde{\Omega}=\Omega\sqrt{T_1T_2}$ and dot denotes $\partial/\partial (t-z/c)$.
 As seen from the Fig. 3, there are substantial differences in the propagation of pulses
 in two level and three level media. Note that the time $T_1$ is
 equal to $1/2\Gamma_2$. We believe that, in the light of the
 energy level diagram of Ruby, it is more appropriate to model it
 as a three level system.

 We next consider input pulse as a modulated
 pulse given in Eq. (5). The output pulse is modulated with a
 phase shift (time delay). We show this time delay as a function
 of modulation frequency for two different pump powers in the Fig. 4. The
 results
 in Figs. 2 - 4 are in excellent agreement with the experimental
 data({\it cf} for example Fig. 3 of Ref.\cite{Bigelow} with our Fig. 4).

 {\it Media with reverse saturation }----For the superluminal propagation in Alexandrite,
 Bigelow {\it et al.}  recognized how the reverse saturation
 mechanism \cite{Malcuit} can be at work in a material like
 BeAl$_2$O$_4$ doped with Cr$^{3+}$ ions
 and with some   Cr$^{3+}$ ions  replaced by Al$^{3+}$.
 The reverse saturation produces an antihole in the susceptibility for
 the probe in presence of a pump field. The antihole  can result
 in the superluminal propagation. In what follows we show how the
 measurement can follow by modelling the system as a four level
 system to account for reverse absorption. The model is shown in the Fig.
 5,
 where state $^4A_2$ as $|g\rangle$, the absorption bands $^4T_2$ and $^4T_1$ as
 $|e_1\rangle$ and the  level $^2E$ as $|e_2\rangle$.
 The density matrix equations are now given by
 \begin{eqnarray}
 \dot{\rho}_{_{_{gg}}} &=& 2\Gamma_2 \rho_{_{_{22}}} + i
 \Omega(\rho_{_{_{1g}}}-\rho_{_{_{g1}}})\nonumber\\
 \dot{\rho}_{_{_{22}}} &=&
 2\Gamma_1\rho_{_{_{11}}}
 -2\Gamma_2\rho_{_{_{22}}}+2\Gamma_3\rho_{_{_{33}}}\nonumber\\
 \dot{\rho}_{_{_{33}}} &=& -2\Gamma_3 \rho_{_{_{33}}} + i
 \Omega(\rho_{_{_{23}}}-\rho_{_{_{32}}})\\
 \dot{\rho}_{_{_{32}}} &=& -\Gamma_3\rho_{_{_{32}}} + i
 \Omega(\rho_{_{_{22}}}-\rho_{_{_{33}}})\nonumber\\
 \dot{\rho}_{_{_{1g}}} &=& -\Gamma_1\rho_{_{_{1g}}} + i
 \Omega(\rho_{_{gg}}-\rho_{_{_{11}}})\nonumber\\
 \rho_{_{gg}}&+&\rho_{_{_{11}}}+\rho_{_{_{22}}}+\rho_{_{_{33}}}=1.\nonumber
 \end{eqnarray}
 Following the same procedure as in the case of
 Ruby, we have derived the working equations
 \begin{eqnarray}
 \frac{\dot{\rho}_{_{gg}}}{2\Gamma_2}
 &=&(1-\rho_{_{_{gg}}})-2\tilde{\Omega}^2\rho_{_{_{gg}}}\\
 \frac{\partial \tilde{\Omega}}{\partial z}
 &=& -
 \frac{\alpha_0}{2}\tilde{\Omega}\rho_{_{_{gg}}}-\frac{\tilde{\alpha_0}}{2}
 \tilde{\Omega}(1-\rho_{_{_{gg}}}),
 \end{eqnarray}
 where $\tilde{\alpha}_0$ gives the reverse saturation. Following
 the experimental data of Bigelow {\it et al.}\cite{Boyd}, we estimate
 $(\tilde{\alpha}_0/\alpha_0) \thickapprox 4 $ . The Eqs. (10) and (11)
 are numerically integrated for the input Gaussian pulse given by Eq. (4).
  A representative set of results is shown in the Fig. 6.
 This Figure
 also shows how the group velocity and net transmission depends on
 the peak intensity of the Gaussian pulses. It should be borne in
 mind that in the range of the intensities of  Fig. 6, no
 perturbation theory can be used. One has to study the full
 nonlinear behavior. We also notice that the input pulses get
 distorted in shape. The distortion becomes more pronounced as the
 nonlinearity of the medium becomes more pronounced.

 In conclusion, we have shown how to model the propagation of intense
 pulse in solid state media with very strong relaxation effects.
 The media can exhibit either saturated absorption or reverse
 absorption.Our modelling goes well beyond the traditional
 pump-probe approach. We specifically present results on
 the propagation of pulses in Ruby and Alexandrite.
 Our model would also be applicable to other systems
 where reverse absorption could be dominant.

 GSA is grateful to R. Boyd for extensive discussions on
 experiments and the contents of this paper. GSA also thanks E.
 Wolf for partially supporting visit to Rochester through US Air
 Force Office of Scientific Research, grant no. F49620-03-1-0318
 and the Engineering Research Program of the Office of Basic
 Energy Sciences at the US Department of Energy under grant no.
 DE-FG02-ER 45992.

 \begin{figure}
 \caption{Three level model for Ruby Crystal}
 \end{figure}

 \begin{figure}
 \caption{The solid curve  shows light pulse
 propagating at speed c through 7.25 cm in vacuum. The long
 dashed and dot-dashed curves show light pulses propagating
 through a medium of length 7.25 cm at different input amplitudes.
 The temporal width $\sigma$ of the Gaussian pulse is 20 msec and $1/2\Gamma_2=4.45$ msec.
 The part (b) gives the
 amplitudes, of the output pulse normalized to the input amplitudes.
 The transmission increases with increasing  the input field intensity.}
 \end{figure}

 \begin{figure}
 \caption{Variation of transmissions and group Velocities as
 functions of the input amplitude of the light pulse. The solid
 (dashed) curve gives the intensity transmission of the pulse for
 the two (three) level model of the medium. The corresponding group
 velocities are given by the dotted curve (two level model) and
 the long dashed curve (three level model). The light
 pulse is propagating through the medium of length L=7.25 cm.}
 \end{figure}

 \begin{figure}
 \caption{Time delay of the light pulse as a function of modulation
 frequency for two different input powers. The modulation index,
 m, is equal to 0.05. Note that the output pulse can be fitted to the from
 $I(t)=I_o(t)(1+m\cos(\Delta t+\theta))$. The overall
 transmission is small, e.g., $I_o(t)/I_s=0.00165$ for
 $\Delta/2\Gamma_2=2$}
 \end{figure}

 \begin{figure}
 \caption{Four level model for Alexandrite crystal}
 \end{figure}

 \begin{figure}
 \caption{The solid curve of (a)  shows light pulse
 propagating at speed c through a distance of 7.25 cm in vacuum. The dotted, long
 dashed and dot-dashed curves depict light pulse propagating
 through a medium of length 7.25 cm at different input amplitudes.
 The pulse width $\sigma$ is 500 $\mu$sec, whereas $1/2\Gamma_2$=250$\mu$sec. Fig (b) shows the amplitude of
 the output pulse normalized with input amplitude. The transmission is
 decreased on increasing  the input field intensity .}
 \end{figure}


\begin{thebibliography}{9999}



\bibitem{Hau} L. V. Hau, S. E. Harris, Z. Dutton, and C. H. Behroozi, Nature
 (London) {\bf 397}, 594 (1999).
\bibitem{Kasapi} A. Kasapi, M. Jain, G. Y. Yin, and S. E. Harris, Phys. Rev. Lett.
 {\bf 74}, 2447 (1995); O. Schmidt, R. Wynands, Z. Hussein, and D. Meschede, Phys. Rev. A
{\bf 53}, R27 (1996).
\bibitem{Mastko}  A. B. Mastko, O. Kocharovskaya, Y. Rostovtsev, A. S.
 Zibrov, and M. O. Scully, Adv. At., Mol., Opt. Phys. {\bf 46}, 191
 (2001); R. W. Boyd and G. J. Gauthier, in {\it slow and Fast
 Light}, Progress in Optics Vol. 43, edited by E. Wolf (Elsevier
 Amsterdam, 2002).
\bibitem{Kash} M. M. Kash, V. A. Sautenkov, A. S. Zibrov, L. Hollberg, G. R.
 Welch, M. D. Lukin, Y. Rostovtsev, E. S. Fry, and M. O. Scully, Phys. Rev. Lett.
 {\bf 82}, 5229 (1999).
 \bibitem{Bud} D. Budker, D. F. Kimball, S. M. Rochester, and V. V. Yashchuk,
 Phys. Rev. Lett. {\bf 83}, 1767 (1999).
\bibitem{Turukhin} A. V. Turukhin, V. S. Sudarshanam, M. S. Shahriar, J. A.
 Musser, B. S. Ham and P. R. Hemmer, Phys. Rev. Lett. {\bf 88}, 023602 (2002).
\bibitem{Liu} C. Liu, Z. Dutton, C. H. Behroozi, and L. V. Hau,
 Nature (London) {\bf 409}, 490 (2001).
\bibitem{Phillips}  M. Fleischhauer and M. D. Lukin, Phys. Rev. Lett. {\bf
 84}, 5094 (2000); M. Fleischhauer and M. D. Lukin, Phys. Rev. A {\bf 65}, 022314
 (2002); D. F. Phillips, A. Fleischhauer, A. Mair, R. L. Walsworth, and
 M. D. Lukin, Phys. Rev. Lett. {\bf 86}, 783
 (2001).
\bibitem{Tarak} T. N. Dey and G. S. Agarwal, Phys. Rev. A {\bf
67}, 033813 (2003).
\bibitem{Grobe} R. Grobe, F. T. Hioe, and J. H. Eberly, Phys. Rev. Lett. {\bf
 73}, 3183 (1994).
\bibitem{Cerboneschi} E. Cerboneschi, F. Renzoni, and E. Arimondo, J. Opt. B
 {\bf 4}, S267 (2002).
\bibitem{Eberly}J. H. Eberly and V. V. Kozlov, Phys. Rev. Lett. {\bf 88},
 243604 (2002); A. Rahman and J. H. Eberly, Phys. Rev. A {\bf 58}, R805 (1998).
 \bibitem{Juzeliunas} G. Juzeliunas, and H. J. Carmichael, Phys. Rev. A {\bf 65},
 021601(R) (2002).
 \bibitem{Mewes}C. Mewes and M. Fleischhauer, Phys. Rev. A {\bf
 66}, 033820 (2002).
\bibitem{Bigelow} M. S. Bigelow, N. N. Lepeshkin, and R. W. Boyd, Phys.
Rev. Lett. {\bf 90}, 113903 (2003).
\bibitem{Harris}S. E. Harris, Physics Today, {\bf 50}(7), 36 (1997).
\bibitem{Hillman} L. W. Hillman, R. W. Boyd, J. Krasinski, and C. R. Stroud, Jr.
, Opt. Commun. {\bf 45}, 416 (1983).
\bibitem{Boyd} M. S. Bigelow, N. N. Lepeshkin, and R. W. Boyd,
Science, {\bf 301}, 200 (2003).
\bibitem{Chu} S. Chu and S. Wong, Phys. Rev. Lett. {\bf 48}, 738
(1982).
\bibitem{Wang} L. J. Wang, A. Kuzmich, and A. Dogariu, Nature (London) {\bf
406}, 277 (2000); A. Dogariu, A.Kuzmich, and L. J. Wang, Phys.
Rev. A {\bf 63}, 053806 (2001).
\bibitem{Chiao} For theoretical work on superluminal propagation
see R. Y. Chiao, Phys. Rev. A {\bf 48}, R34 (1993); E. L. Bolda,
R. Y. Chiao, and J. C. Garrison, Phys. Rev. A {\bf 48}, 3890
(1993); G. S. Agarwal, T. N. Dey, and S. Menon, Phys. Rev. A {\bf
64}, 053809 (2001); D. Bortman-Arbiv, A. D. Wilson-Gordon, and H.
Friedmann, Phys. Rev. A {\bf 63}, 043818 (2001).
\bibitem{Allen}L. Allen and J. H. Eberly, {\it Optical Resonance
and Two Level Atoms}~(Dover, New York, 1987), p. 90.
\bibitem{Mollow}B. R. Mollow, Phys. Rev. A {\bf 5}, 2217 (1972).
\bibitem{Malcuit} M. S. Malcuit, R. W. Boyd, L. W. Hillman, and C.
R. Stroud, Jr., JOSA B {\bf 1} 73, (1984); K. V. Yumashev, N. V.
Kuleshov, P. V. Prokoshin, A. M. Malyarevich, and V. P. Mikhailov,
Appl. Phys. Lett. {\bf 70}, 2523 (1997); Z. Burshtein, P. Blau, Y.
Kalisky, Y. Shimony, and M. R. Kokta, IEEE J. Quantum Electron,
{\bf 34}, 292 (1998).

 \end{thebibliography}
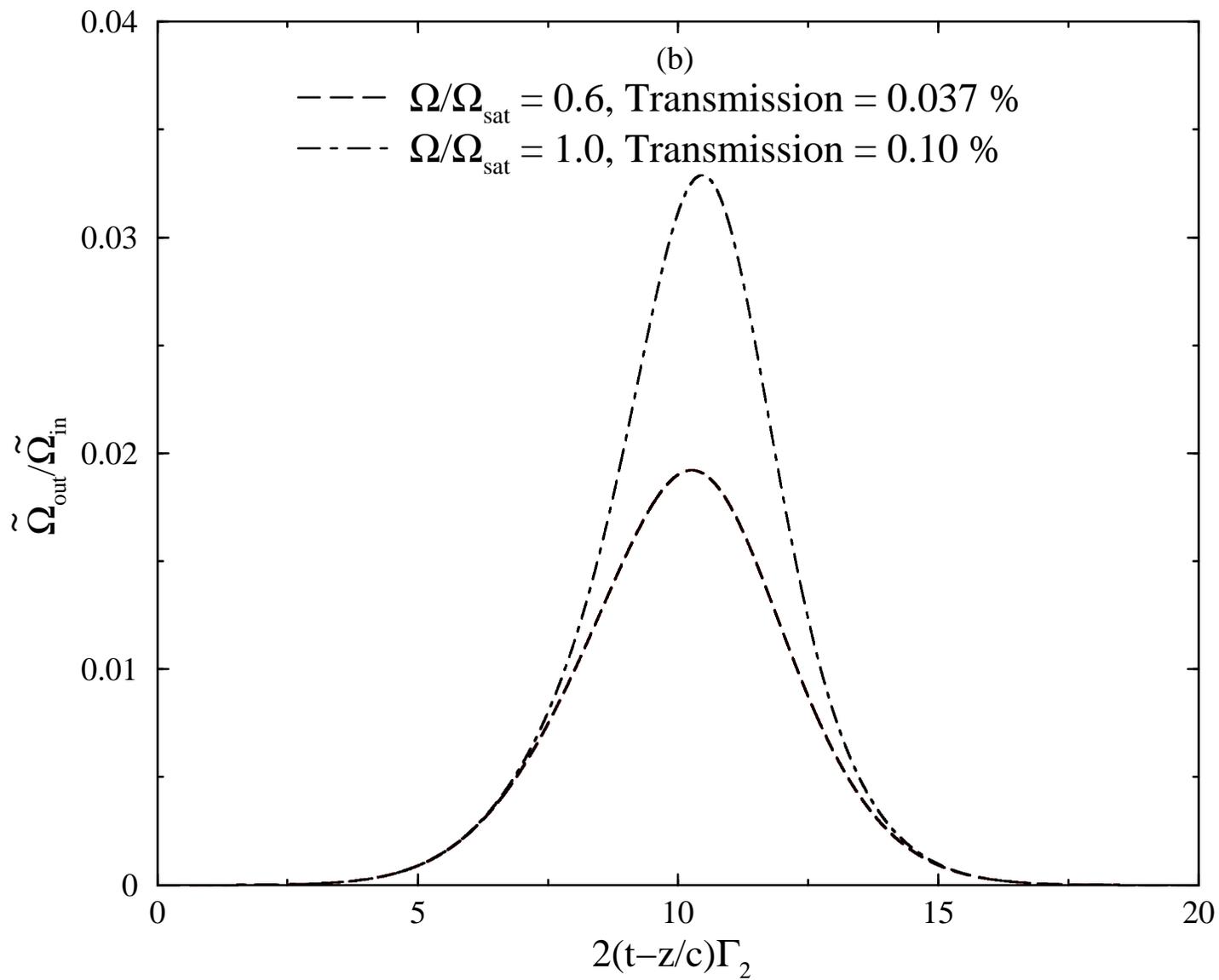
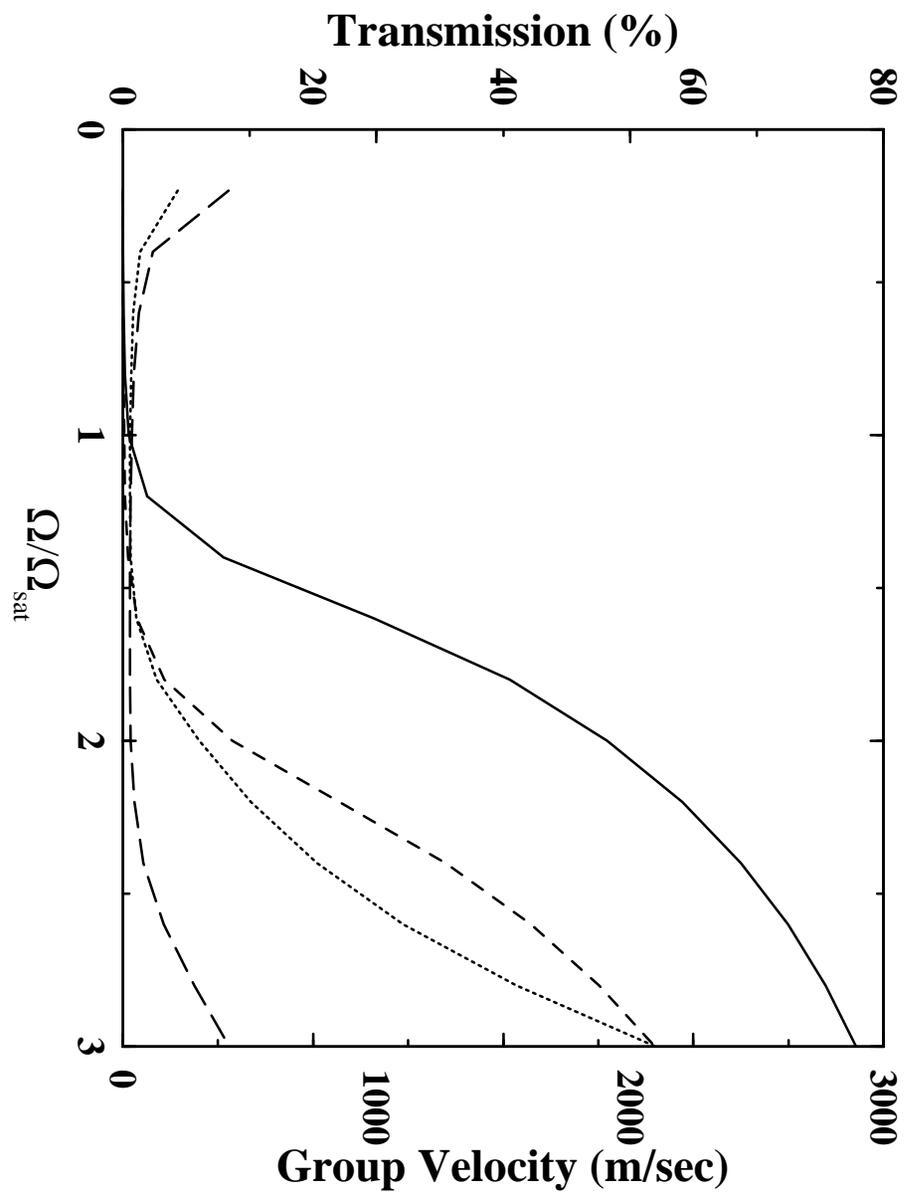
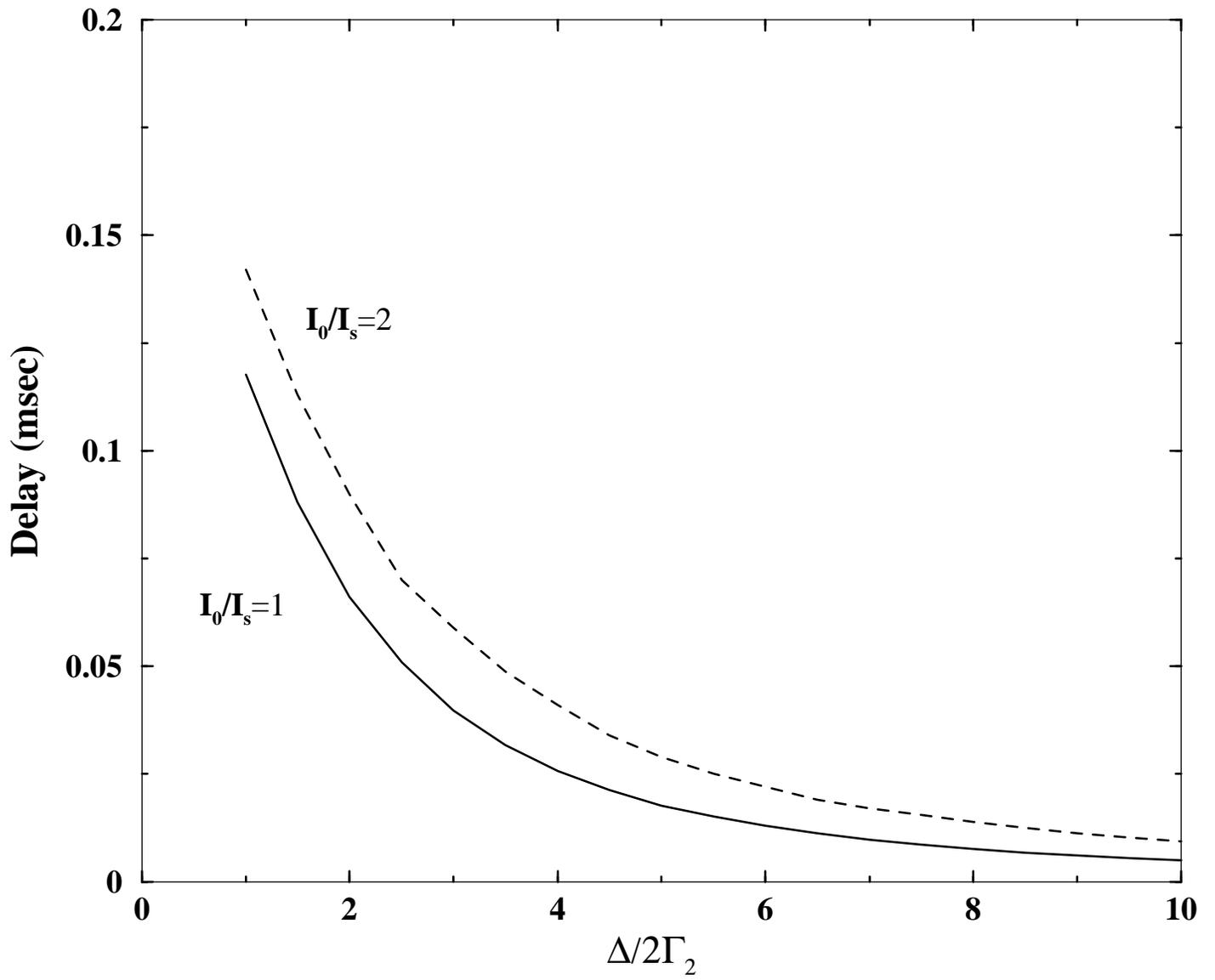
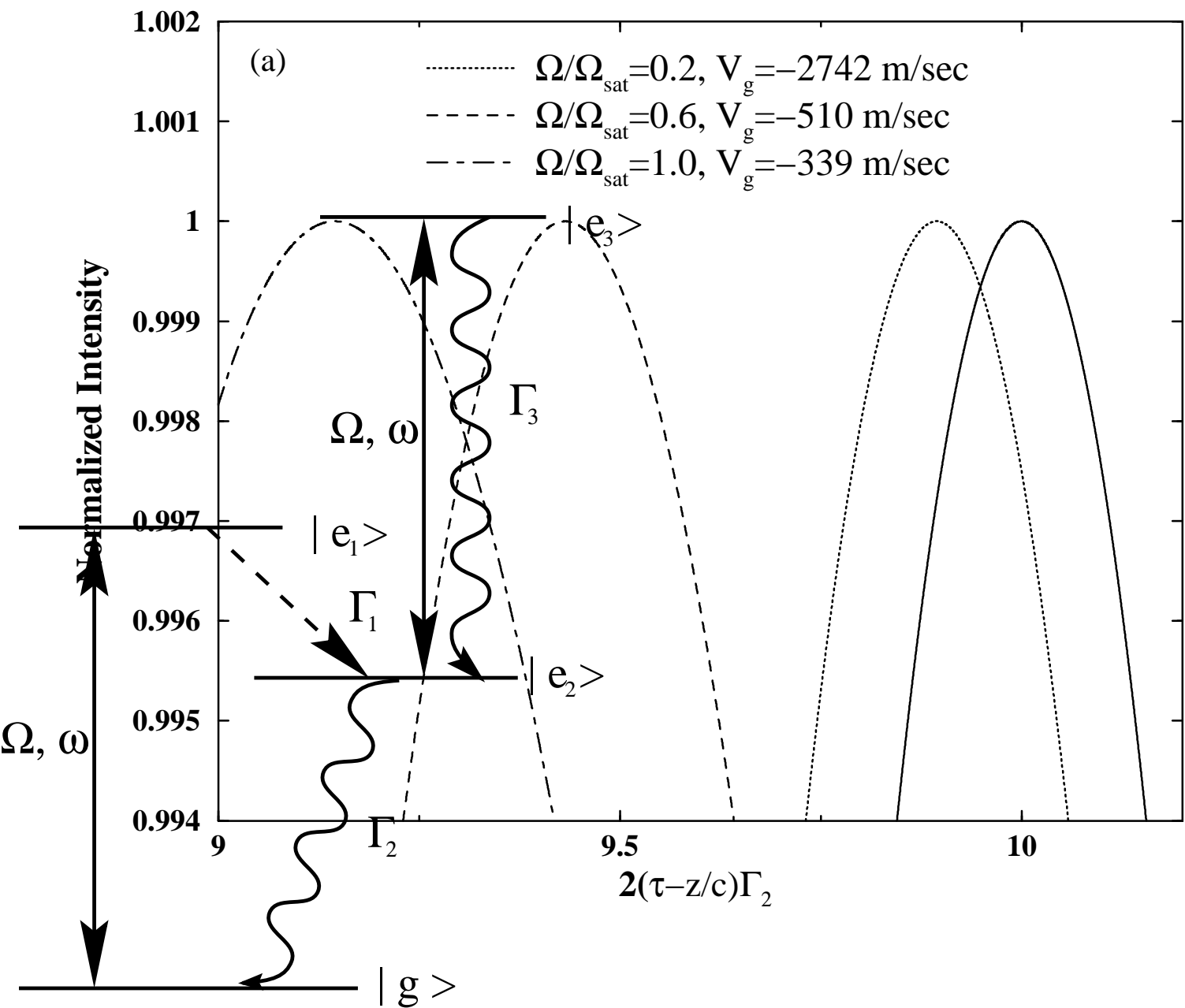
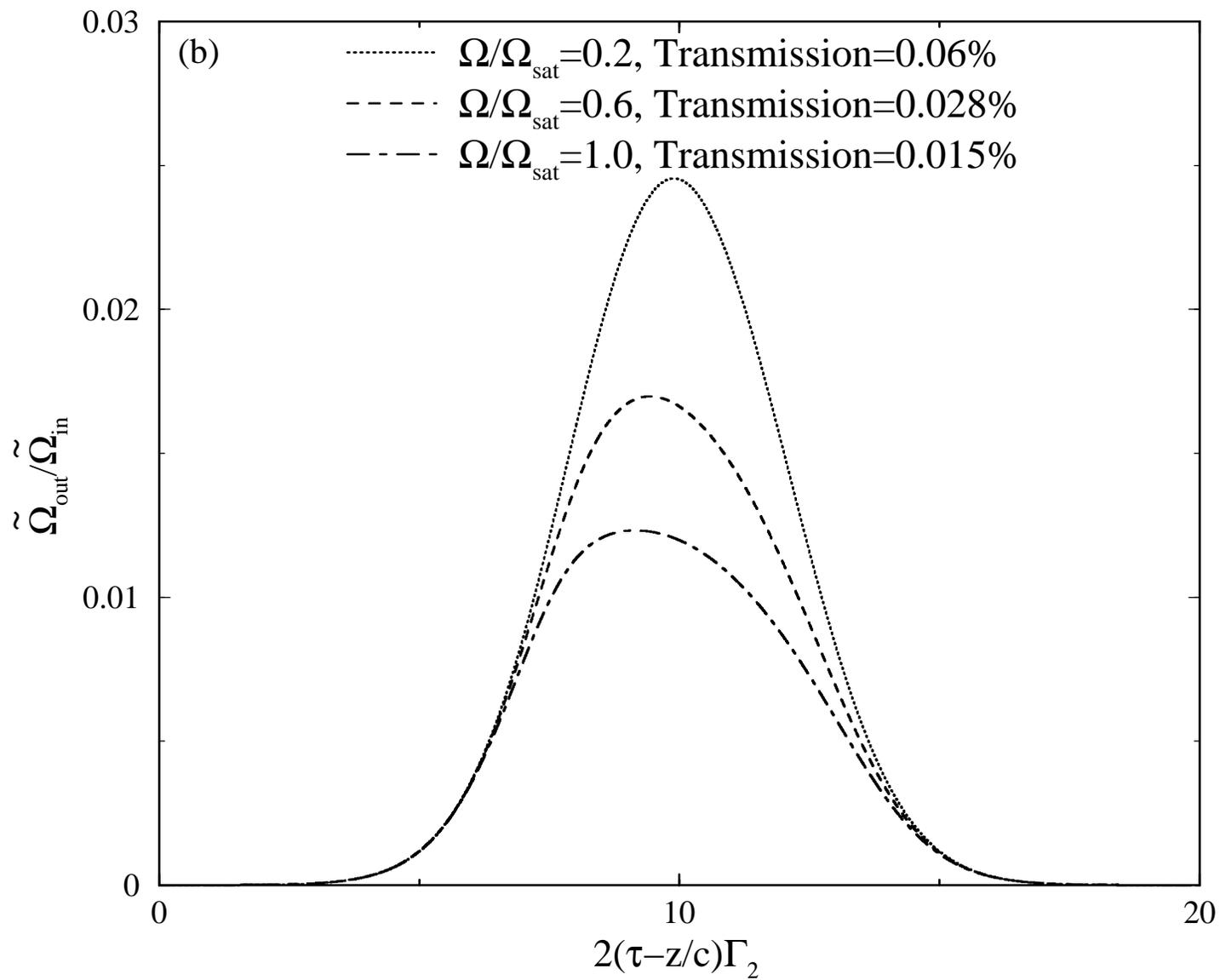
\end{document}